\documentclass[prl,superscriptaddress,twocolumn,noshowpacs]{revtex4-1}
\usepackage{graphicx,amsmath,amssymb,amsfonts,dsfont,natbib,physics}
\usepackage{multirow}
\usepackage{lipsum,color}
\usepackage{verbatim}
\usepackage{commath}
\usepackage{bm}
\usepackage{float}
\usepackage{hyperref}

\begin{document}

\title{Lattice supersymmetry and order-disorder coexistence in the tricritical Ising model}

\author{Edward O'Brien}
\affiliation{Rudolf Peierls Centre for Theoretical Physics, 1 Keble Road, Oxford, OX1 3NP, United Kingdom}
\author{Paul Fendley}
\affiliation{Rudolf Peierls Centre for Theoretical Physics, 1 Keble Road, Oxford, OX1 3NP, United Kingdom}
\affiliation{All Souls College, Oxford, OX1 4AL, United Kingdom}

\vskip1in
\date{\today} 

\begin{abstract} 
We introduce and analyze a quantum spin/Majorana chain with a tricritical Ising point separating a critical phase from a gapped phase with order-disorder coexistence. We show that supersymmetry is not only an emergent property of the scaling limit, but manifests itself on the lattice. Namely, we find explicit lattice expressions for the supersymmetry generators and currents. Writing the Hamiltonian in terms of these generators allows us to find the ground states exactly at a frustration-free coupling. These confirm the coexistence between two (topologically) ordered ground states and a disordered one in the gapped phase. Deforming the model by including explicit chiral symmetry breaking, we find the phases persist up to an unusual chiral phase transition where the supersymmetry becomes exact even on the lattice.

\end{abstract} 

\maketitle

\paragraph{Introduction}

The tricritical Ising model in two dimensions exhibits a host of remarkable features.
Some known since the '70s include a line of phase coexistence, a tricritical point where the line ends, and a non-trivial renormalization-group (RG) flow from it to the Ising critical point \cite{Blume1971}. Some understood in the '80s include a description of the scaling limit of the tricritical point in terms of a conformal field theory (CFT). Even more strikingly, this strongly interacting CFT possesses supersymmetry,  somewhat optimistically viewed as the first such manifestation in nature \cite{Friedan1984b}.  
The supersymmetry is unbroken in the scaling limit of the phase coexistence line, while it is spontaneously broken along the flow to Ising. A marvelous consequence of the latter is that the ensuing Goldstinos are precisely the Majorana fermions of the Ising CFT \cite{Kastor1988}. The corresponding field theories are integrable and their quasiparticle excitations and their scattering matrices can be found exactly \cite{Zamolodchikov1989,Zamolodchikov1991a,Zamolodchikov1991b,Schoutens1990}.

A host of lattice models reproduces the universal physics of the transition line. The classic example is the Blume-Capel model, best thought of as the Ising model with vacant sites allowed. Numerics and an RG analysis indicate when the fugacity for the vacancies is large enough, the usual order-disorder Ising transition line indeed converts to a first-order line \cite{Blume1971}. The lines meet at a non-trivial  tricritical point. However, couplings must be fine-tuned to find the transition line, as no lattice symmetry protects it.  In an integrable model in the same universality class, interacting hard squares \cite{Baxter1982,Huse1984}, the coexistence line can be located precisely, and the vanishing of the bulk ground-state energy characteristic of a supersymmetric theory derived exactly. Again, however, fine-tuning is required to obtain the desired physics. 

Motivated both by the presence of supersymmetry and by the interesting topological features of Majorana fermions, several groups recently revisited this story \cite{Grover2014,Rahmani2015a,Rahmani2015b}. They introduced quantum spin chains with a self-duality that fixes the transition line exactly, and gave convincing numerical checks (extended further in \cite{Zou2017}) that the models display the expected universal physics.

However, in these lattice models the supersymmetry is seen only indirectly, as an emergent property of the scaling limit, i.e.\ only appears in the field-theory description valid at and near the tricritical point. Since a number of spin chains with explicit supersymmetry become supersymmetric field theories \cite{Fendley2002,Fendley2003,Huijse2011,Bauer2012} in the scaling limit, it is rather curious that no such model for the tricritical Ising model was known. The purpose of this paper is to introduce and analyze a quantum spin chain in the same univerality class, but where the connection to supersymmetry is manifest on the lattice.

Our chain retains the desirable characteristics of earlier models, and includes several new ones. Its self-duality means no fine tuning is needed to find the phase where the effective field theory is supersymmetric. It describes phase coexistence of order with disorder, or, in the fermionic picture \cite{Kitaev2001}, coexistence of topologically ordered ground states that admit fermionic edge zero modes with a non-topological ground state that does not. This type of coexistence occurs without including extra degrees of freedom such as the vacancies in Blume-Capel or a second Majorana chain in Ref.\ \onlinecite{Grover2014}. Moreover, our Hamiltonian has the tricritical point at couplings of order 1, whereas that of Ref.\ \onlinecite{Rahmani2015a} requires the interaction terms to be $\sim 250$ times larger than the non-interacting ones. Yet another advantage of our Hamiltonian is that it exhibits a ``frustration-free'' point, where the exact ground states can be found explicitly. These make the order-disorder coexistence precise and rigorous.  

The most prominent new feature of our model is that the Hamiltonian can be written as a sum of the squares of two fermionic operators, just like a supersymmetric field theory familiar from particle physics. We use them to find lattice analogs of the supersymmetry generators, i.e.\ lattice operators that in the scaling limit have the same properties as the corresponding fields. This form allows for a chiral deformation of the model to a line of couplings where supersymmetry becomes exact even on the lattice. We show that this line describes an unusual chiral phase transition out of the supersymmetric phase.  


\paragraph{The model}

We study throughout the Ising/Majorana quantum spin chain with additional interactions. The Hilbert space is comprised of $L$ two-state systems, with Hermitian operators acting on it given by $\sigma^r_j=1\otimes 1 \cdots 1 \otimes \sigma^r \otimes 1\cdots 1$, with $\sigma^r$ the Pauli matrix with $r=x,y,z$ acting on the two-state system at the $j$th site. An extremely useful basis of operators is given by the Jordan-Wigner transformation to Majorana fermions
\begin{equation}
\gamma_{2j-1}=\sigma_j^z\prod\limits_{k=1}^{j-1}\sigma_k^x\ , \qquad
\gamma_{2j}= -i \sigma^x_j \gamma_{2j-1}\ .
\end{equation}
These Majorana operators are Hermitian and obey $\{\gamma_a,\gamma_{a'}\}=2\delta_{a,a'}$.  We consider only ${\mathbb Z}_2$-symmetric Hamiltonians commuting with the spin-flip (or fermion-number parity) operator $\mathcal{F}=\prod_j\sigma_j^x$. The simplest one is that of the Ising chain at its self-dual critical point:\begin{equation}
H_I=-\sum\limits_j\left(\sigma_j^x+\sigma_j^z\sigma_{j+1}^z\right)
=i\sum\limits_a\gamma_a\gamma_{a+1}\ ,
\end{equation}
where we ignore subtleties with the boundary conditions. Roughly speaking, self-duality amounts to invariance of $H$ under the shift $\gamma_a\to\gamma_{a+1}$; a more thorough discussion can be found in \cite{Aasen2016}. The scaling limit of this Hamiltonian is given by the Ising CFT \cite{Belavin84}.

To give a direct connection to supersymmetry, we add to the Hamiltonian the three-spin interaction 
\begin{align}
H_3&=\sum\limits_j\left(\sigma_j^x\sigma_{j+1}^z\sigma_{j+2}^z+\sigma_j^z\sigma_{j+1}^z\sigma_{j+2}^x\right)\cr
&=-\sum\limits_a\gamma_{a-2}\gamma_{a-1}\gamma_{a+1}\gamma_{a+2}\ .
\end{align}
The Hamiltonian then can be rewritten as the sum of the squares of two fermionic Hermitian operators:
\begin{align}
H=  (Q^+)^2 + (Q^-)^2= 2\lambda_IH_I+\lambda_3H_3 + E_0\ ,
\label{Hdef}
\end{align}
where $\lambda_3\ge 0$, $E_0=L(\lambda_I^2+\lambda_3^2)/\lambda_3$, and
\begin{align}
Q^\pm =\frac{1}{2\sqrt{\lambda_3}}\sum\limits_a(\pm 1)^a\left(\lambda_I\,\gamma_a\pm i\lambda_3\,\gamma_{a-1}\gamma_a\gamma_{a+1}\right)\ .
\label{Qdef}
\end{align}
Under duality, $Q^\pm \to \pm Q^\pm$, so $H$ remains self-dual.
Since $Q^+$ and $Q^-$ do not commute, supersymmetry here is not exact on the lattice. However, we show below that not only are these operators explicit lattice analogs of the supersymmetry generators, but also lead to analogs of the supersymmetry currents.

The interaction $H_3$ is an irrelevant perturbation of the Ising fixed point,  as there are no relevant self-dual operators in the Ising CFT. The least-irrelevant self-dual operator there is $T\overline{T}$, the product of left- and right-moving components of the energy-momentum tensor \cite{Kastor1988}.
The four-fermi terms in $H_3$ differ from those ($\gamma_{a-2}\gamma_{a-1}\gamma_{a}\gamma_{a+1}$) analyzed in \cite{Rahmani2015a,Rahmani2015b}, but both are expected to renormalize onto  $T\overline{T}$ in the Ising CFT. Both perturbations thus should lead to the same effective field theory. The Hamiltonian with both $H_I$ and $H_3$ present is not integrable, but it is worth noting that $H_3$ by itself is not only integrable and supersymmetric, but has a hidden free-fermion structure \cite{Fendley18}.

\paragraph{Exact ground states with order-disorder coexistence} 

We can find the ground states exactly when $\lambda_I=\lambda_3$. Since all energies found from (\ref{Hdef}) are non-negative, any states annihilated by both $Q^+$ and $Q^-$ are necessarily ground states of $H$ with zero energy.  At $\lambda_I=\lambda_3=1$,  $Q^\pm$ each can be rewritten in two useful ways \cite{Sannomiya2017b}:
\begin{align}
Q^{\pm}\Big|_{\lambda_3=\lambda_I}&=\sum\limits_j\left(\gamma_{2j-2}\pm\gamma_{2j+1}\right)\left(\frac{1}{2}+\frac{i}{2}\gamma_{2j-1}\gamma_{2j}\right) \cr
& =\sum\limits_j\left(\gamma_{2j+2}\pm \gamma_{2j-1}\right)\left(\frac{1}{2}+\frac{i}{2}\gamma_{2j}\gamma_{2j+1}\right).
\label{Qpm}
\end{align}
Since $\gamma_{2j-1}\gamma_{2j}=i\sigma^x_j$, the first form of \eqref{Qpm} requires that any state where all $\sigma^x_j$ have eigenvalue $1$ is annihilated by both $Q^\pm$ and so is a ground state.
The unique such state is the equal amplitude sum over all basis states $s$:
\begin{align}
\ket{{\cal G}_0}=2^{-L/2}\sum_s \ket{s}\ .
\label{g0}
\end{align}
Likewise, since $\gamma_{2j}\gamma_{2j+1}=i\sigma^z_j\sigma^z_{j+1}$, the second form of \eqref{Qpm} requires that any states where all $\sigma^z_j$ have the same eigenvalue are also ground states. Letting $\sigma^z\ket{\uparrow}= \ket{\uparrow}$ and $\sigma^z\ket{\downarrow}= -\ket{\downarrow}$, these ground states are
\begin{align}
\ket{{\cal G}_\uparrow} = \ket{\uparrow\uparrow\uparrow\cdots\uparrow},\qquad \ket{{\cal G}_\downarrow} = \ket{\downarrow\downarrow\downarrow\cdots\downarrow}.
\label{gpm}
\end{align}
We prove in the appendix that these three are the only ground states for $\lambda_I=\lambda_3$. The lowest-lying excitations are comprised of domain walls between the ground states, in accordance with exact field-theory results \cite{Zamolodchikov1989,Zamolodchikov1991a}. 

The ground state $\ket{{\cal G}_0}$ is completely disordered, while the perfectly ordered states $\ket{{\cal G}_\uparrow}$ and $\ket{{\cal G}_\downarrow}$ spontaneously break the ${\mathbb Z}_2$ symmetry.  The degeneracy is a consequence of the self-duality: if any of these states is a ground state, they all must be. Indeed, under duality, $\ket{{\cal G}_\downarrow}\to\sqrt{2}\ket{{\cal G}_0}$ and $\ket{{\cal G}_\uparrow}\to\sqrt{2}\ket{{\cal G}_0}$, while $\ket{{\cal G}_0}\to\ket{{\cal G}_\downarrow}+\ket{{\cal G}_\uparrow}$, as shown in Ref.\ \onlinecite{Aasen2016}. Upon breaking the self-duality, this coexistence line becomes a first-order transition separating ordered and disordered phases.


\paragraph{The phase diagram}

At the frustration-free coupling $\lambda_I=\lambda_3$, our Hamiltonian (\ref{Hdef}) displays the physics of the gapped self-dual phase, with one ${\mathbb Z}_2$ symmetry-preserving ground state coexisting with two symmetry-broken ones. Doing perturbation theory around it gives an energy splitting proportional to $|\lambda_I-\lambda_3|^{L/2}$, and so exponentially small. The order-disorder coexistence therefore persists throughout a gapped phase, as expected \cite{Blume1971,Huse1984}. The frustration-free point is presumably the basin for the RG flows in this phase. An additional piece of evidence for such is that the exact solution \cite{Fendley18} of the gapless pure four-fermi point $\lambda_I=0$ indicates that the dynamical critical exponent there is $z=3/2$. This suggests the emergent free fermions there are of dimension $1/4$ and that a fermion bilinear like the Ising perturbation is relevant, so all flows starting at $\lambda_3>\lambda_I$ flow to the basin.

\begin{figure}[ht]
\includegraphics[width=\linewidth]{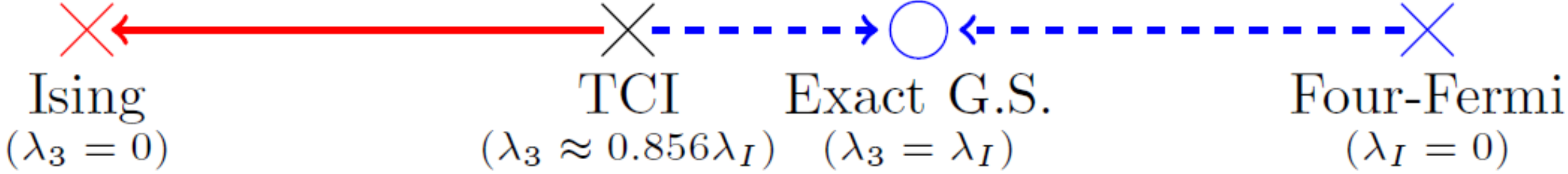}
\vspace{-0.3cm}\caption{Phase diagram and RG flows of (\ref{Hdef}).\vspace{-0.2cm}}  
\label{fig:Phase_diagram}
\end{figure}

Because the Ising critical point is also stable under RG flows along the self-dual line, an unstable tricritical point must separate the critical Ising phase from the gapped coexistence phase at some value of $\lambda_3$ less than $\lambda_I$. Thus our results yield the phase diagram in fig.\ \ref{fig:Phase_diagram}, precisely that expected along the self-dual line. 

To confirm this picture and locate the tricritical point, we do extensive numerics. 
A key tool is that ratios of energy differences are universal at a critical point described by a CFT \cite{Blote1986,Affleck1986}: $(E_{\alpha}-E_{\beta})/(E_{\gamma}-E_{\delta})=(\Delta_{\alpha}-\Delta_{\beta})/(\Delta_{\gamma}-\Delta_{\delta})$, where $\Delta_\alpha$ is the dimension of the CFT operator that creates an excited state with energy $E_\alpha$. Since the spin-flip operator $\mathcal{F}$ commutes with $H$ and obeys $\mathcal{F}^2=1$, we split the Hilbert space into sectors with eigenvalues $\pm 1$ of ${\mathcal F}$. We label the corresponding energies of $H$ with periodic and anti-periodic boundary conditions as $P^\pm_r$ and $A_r^\pm$ respectively with $r=0,1,\dots$ (anti-periodic means we flip the sign of all terms in $H$ involving $\sigma^z_L\sigma^z_{1}$). Some of the CFT values (see e.g.\ \cite{Friedan1984b,Rahmani2015a}) and the central charge $c$ are:
 \begin{table}[h]
\begin{ruledtabular}
\begin{tabular}{ccccc}
CFT & $c$ & $R_1=\frac{A_0^- -P_0^+}{P_1^+-P_0^+}$ & $R_2=\frac{P_0^--P_0^+}{P_1^+-P_0^+}$ & $R_3=\frac{P_1^- -P_0^+}{P_1^+-P_0^+}$\\
\hline
Ising & $\frac{1}{2}$ & $\frac{1}{2}$ & $\frac{1}{8}$ & $\frac{9}{8}$\\
TCI & $\frac{7}{10}$ & $\frac{7}{2}$ & $\frac{3}{8}$ & $\frac{35}{8}$\\
\end{tabular}
\end{ruledtabular}
\end{table}

\begin{figure}[t]
\includegraphics[width=\linewidth]{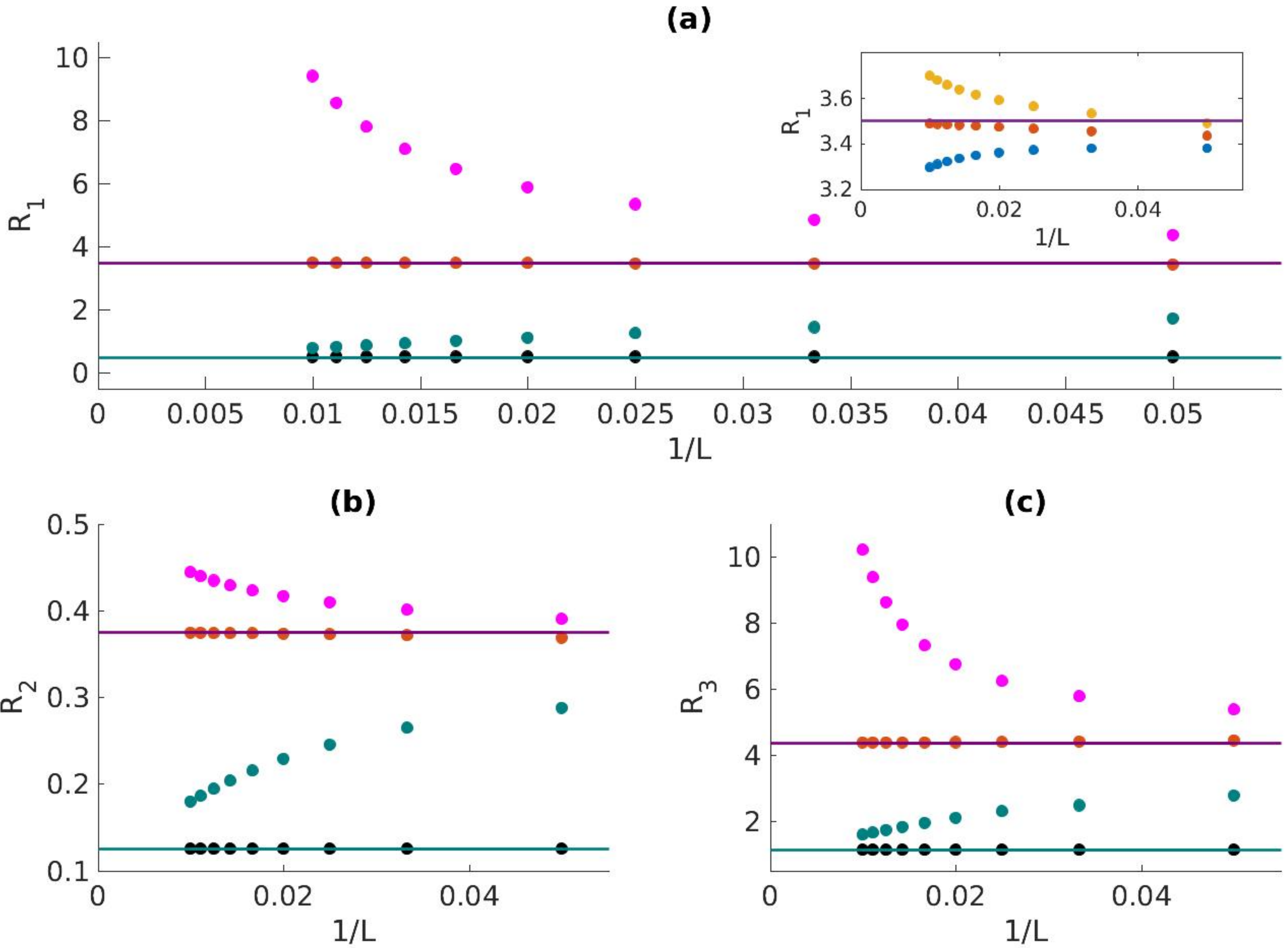}
\vspace{-0.5cm}\caption{The ratios $R_1$,$R_2$ and $R_3$ from the DMRG. Black, green, red and mauve points are  $\lambda_3/\lambda_I=0,0.8,0.856,0.87$ respectively, while blue and yellow in the inset are $0.855$ and $0.857$.  The green and purple lines denote the theoretical values for the Ising and TCI CFTs respectively. \vspace{-0.5cm}}
\label{fig:ratios}
\end{figure}

We computed these ratios numerically using the DMRG with the ITensor C++ library \cite{iTensor}. Our results are given in figure \ref{fig:ratios}. For $\lambda_3/\lambda_I \lessapprox 0.856$ the ratios tend to the values predicted by the Ising CFT as $L$ is increased, while at $\lambda_3/\lambda_I\approx 0.856$ the ratios converge very nicely to the values predicted by the TCI CFT. We also computed the central charges from the coefficient of the log term in the entanglement entropy \cite{Calabrese2004}, and found the appropriate values $c=1/2$ and $c=7/10$ respectively. We thus confirm that the non-trivial critical point in $H_3$ is indeed in the tricritical Ising universality class. We checked numerically that as $\lambda_3/\lambda_I$ is increased to the tricritical point, the bulk ground-state energy decreases as $|\lambda_3-.856\lambda_I |^\nu$,  with $\nu\approx 5/2$, as predicted by perturbed CFT \cite{Kastor1988}. For $\lambda_3> .856 \lambda_I$, we found the differences between  $P_0^+$, $P_1^+$ and $P_0^-$ all go to zero exponentially fast as $L$ is increased, while all other differences indicate a gap. We thus confirm there are three ground states throughout a gapped phase with order-disorder coexistence. In the appendix we consider the effect of adding other couplings expected to be similarly relevant to $H_3$ and see that the tricritical transition survives.

\paragraph{Supersymmetry currents on the lattice}

We have given strong evidence that the scaling limit of $H$ is indeed the supersymmetric field theory of \cite{Kastor1988,Zamolodchikov1989,Zamolodchikov1991a}. We now turn to direct manifestations of supersymmetry on the lattice.  The Hamiltonian of a 1+1-dimensional supersymmetric field theory is
\begin{align}
H_{\rm field\ theory}= \left(\int dx\,G\right)^2 + \left(\int dx\,\overline{G}\right)^2\ .
\end{align}
where the integrals are over space with $G$ and $\overline{G}$ the components of the supersymmetry current in complex coordinates. 
Comparing this Hamiltonian to our lattice one (\ref{Hdef},\ref{Qdef}) makes rather obvious the identification $Q^+ \to \int G$, $\ Q^- \to \int\overline{G}$ in the scaling limit. A further check is that along the phase coexistence line,  $Q^\pm$ act on the domain walls just as the supersymmetry generators do in the corresponding integrable field theory \cite{Zamolodchikov1989,Schoutens1990}. 

We go further to find lattice analogs of the supersymmetry currents themselves, i.e.\  define lattice fermionic operators $G_j$ whose correlation functions become those of $G$ in the continuum limit, while keeping $Q^+=\sum_jG_j$. 
Examining (\ref{Qdef}) suggests the simple, but not correct, guess for $G_j$ as $\lambda_I\gamma_{2j}+i\lambda_3 \gamma_{2j-1}\gamma_{2j}\gamma_{2j+1}$.  
The complication is that in the tricritical Ising CFT, there occur other fermionic fields $\psi$ and $\overline{\psi}$ of (left, right) scaling dimensions $(3/5,1/10)$ and $(1/10,3/5)$ respectively. 
In a two-dimensional superconformal field theory, current conservation becomes $\partial_{\overline{z}}G=0=\partial_z\overline{G}$ so that $G(z)$ and $\overline{G}(\overline{z})$ are holomorphic and antiholomorphic fields of dimensions $(3/2,0)$ and $(0,3/2)$ respectively \cite{Friedan1984b}. Since all these fields have the same conformal spin mod 1, lattice operators typically scale to mixtures of the fields, as happens for example to the lattice parafermions in the three-state Potts model \cite{Mong2014}. 

The key to finding the correct lattice analogs is to understand the action of duality on the fields  \cite{Mong2014}.
We can choose conventions so that under duality, $G$, $Q^+$ and $\overline{\psi}$  are even, while $\overline{G}$, $Q^-$ and $\psi$ are odd \cite{Kastor1988,Lassig1991}. Mimicking this behavior with local
operators then suggests 
\begin{align}
G_j&=\lambda_I(\gamma_{2j-1}+\gamma_{2j})+i\lambda_3(\gamma_{2j-2}+\gamma_{2j+1})\gamma_{2j-1}\gamma_{2j},\cr
\overline{G}_j&=\lambda_I(\gamma_{2j-1}-\gamma_{2j})+i\lambda_3(\gamma_{2j+1}-\gamma_{2j-2})\gamma_{2j-1}\gamma_{2j},\cr
\psi_j&= \lambda_I(\gamma_{2j-1}-\gamma_{2j})+i\lambda_3(\gamma_{2j-2}-\gamma_{2j+1})\gamma_{2j-1}\gamma_{2j},\cr
\overline{\psi}_j&=\lambda_I(\gamma_{2j-1}+\gamma_{2j})-i\lambda_3(\gamma_{2j+1}+\gamma_{2j-2})\gamma_{2j-1}\gamma_{2j}.
\label{latticeanalogs}
\end{align}

We checked these identifications by computing the two-point functions at the tricritical point,  although we expect that our identifications hold throughout. 
\begin{figure}[ht]
\includegraphics[width=1\linewidth]{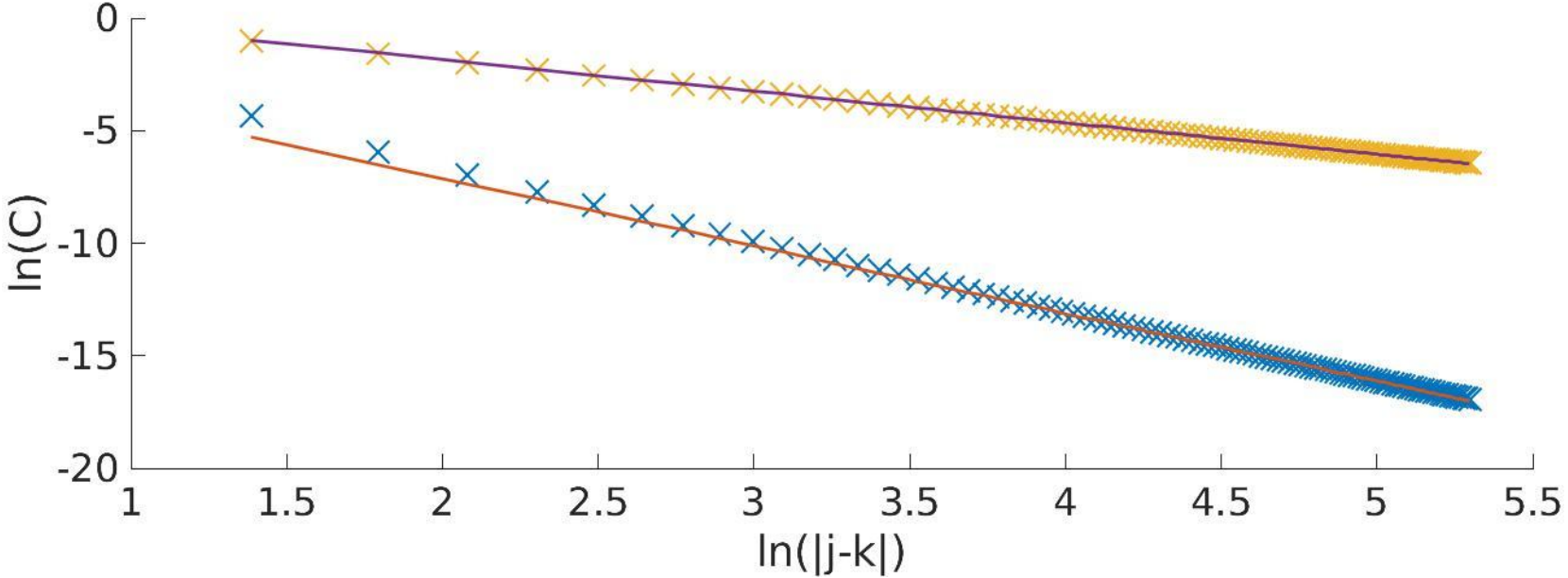}
\vspace{-0.5cm}\caption{Dependence of $C_{jk}=\langle\psi_j\psi_k\rangle$ (yellow crosses) and $\langle G_jG_k\rangle$ (blue crosses) on $|j-k|$  at the tricritical point $\lambda_3/\lambda_I=.856$. The purple and red lines correspond to $A|j-k|^{-1.4}$ and $B|j-k|^{-3}$ respectively, where $A$ and $B$ are fitting parameters. }  
\label{Correlators}
\end{figure} 
The two-point functions of the operators in (\ref{latticeanalogs}) in the ground state at the tricritical point $\lambda_3/\lambda_I=.856$ were computed for $L=400$ and are shown in Fig. \ref{Correlators}. We fitted them to the scaling form $A|j-k|^{-2\Delta_\Phi}$, where $\Delta_{\Phi}$ is the sum of the left and right scaling dimensions. 
As can be seen, the correlators converge very nicely to the expected values of $\Delta_G=\Delta_{\overline G}=3/2$ and  $\Delta_\psi=\Delta_{\overline{\psi}}=7/10$. These numerics thus not only provide another compelling check of the identification of the tricritical point, but that the lattice operators \eqref{latticeanalogs} indeed become the appropriate supersymmetry currents and fermions in the continuum limit.

\paragraph{Explicit lattice supersymmetry in the chiral model}

We now break chiral symmetry by varying the coefficients in front of the $(Q^\pm)^2$ terms, and find that it leads to an unusual phase transition with explicit lattice supersymmetry. Explicitly,
\begin{align}
\widetilde{H}&=
\left(1+\frac{\lambda_c}{\lambda_I}\right)(Q^+)^2 + \left(1-\frac{\lambda_c}{\lambda_I}\right)(Q^-)^2\cr
&= 2\lambda_IH_I+\lambda_3H_3+\lambda_cH_c +E_0\ ,
\label{Hchiral}
\end{align}
where the chiral part of Hamiltonian is 
\begin{equation}
H_c=\sum\limits_j\left(\sigma_j^y\sigma_{j+1}^z-\sigma_j^z\sigma_{j+1}^y\right) 
= -i\sum_a \gamma_{a}\gamma_{a+2}\ .
\label{Hcdef}
\end{equation}

It is straightforward to see that the phases remain intact as long as $|\lambda_c|<\lambda_I$. Along the Ising line, including $H_c$ corresponds to a chiral and marginal perturbation of the CFT by $\int(T-\overline{T})$. This amounts to deforming spacetime by a conformal transformation, and so leaves the model conformally invariant. Indeed, since $[H_c,H_I]=0$, this statement is exact when $\lambda_3=0$. In the gapped phase, the form of \eqref{Hchiral} means that $Q^+$ and $Q^-$ still annihilate the three ground states for $|\lambda_c|<\lambda_I$.  Our numerics confirm this picture, with ground states in the gapped phase at different values of $\lambda_c$ but the same $\lambda_3$ and $\lambda_I$ having overlap $>99\%$. The tricritical transition persists as well, with its location depending very weakly on $\lambda_c$, changing only by about 5\% up to $\lambda_c=\lambda_I$.

\begin{figure}[t]
\includegraphics[width=1\linewidth]{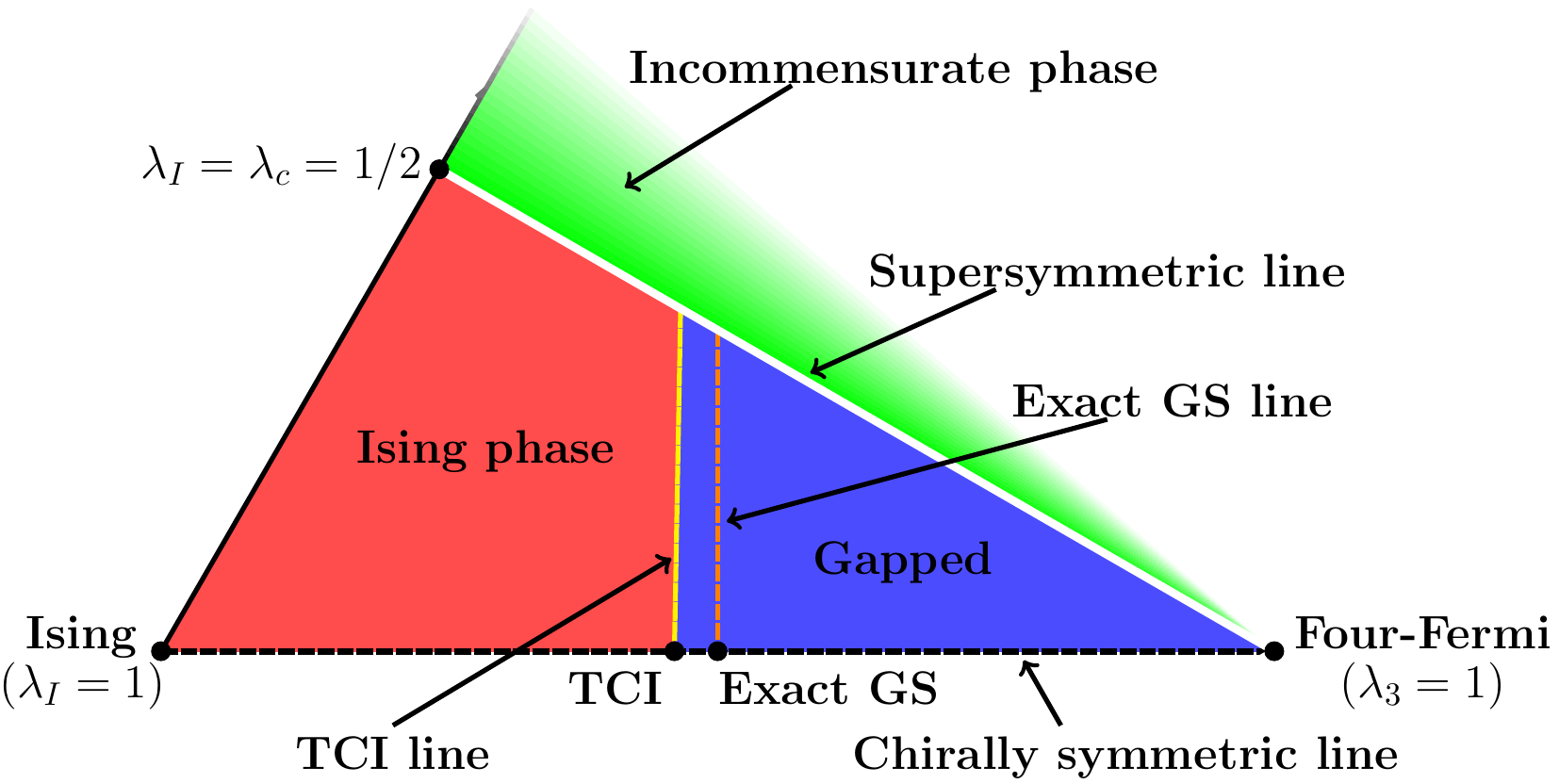}
\vspace{-0.5cm}\caption{Phase diagram including chiral interactions, with $\lambda_I+\lambda_3+\lambda_c=1$.
The diagram is invariant under sending $\lambda_c\to -\lambda_c$. \vspace{-0.5cm}}  
\label{Phase_diagram}
\end{figure} 

For $|\lambda_c|>\lambda_I$ the coefficient of one of the $(Q^\pm)^2$ terms in $\widetilde{H}$ becomes negative. Some states with non-zero momentum become negative in energy, and so we expect phase transitions along the lines $\lambda_c=\pm\lambda_I$ to an incommensurate phase. In the appendix we verify these phase transitions analytically for the Ising line $\lambda_3=0$ and numerically for non-zero $\lambda_3$. 

Along these chiral phase transition lines, the Hamiltonians obey $\widetilde{H}^\pm = 2(Q^\pm)^2$, and so have exact lattice supersymmetry as in Ref.\ \onlinecite{Fendley2002}. Since $[Q^\pm,\widetilde{H}^\pm]=0$ along with $\{ \mathcal{F},\,Q^\pm\}=0$, the spectra in the sectors with eigenvalues $1$ and $-1$ of $\mathcal{F}=1$  are necessarily identical. These Hamiltonians are Majorana analogs \cite{Sannomiya2017b} of the Nicolai model \cite{Nicolai1976}. Along these lines, a rigorous proof \cite{Sannomiya2017b} shows that indeed supersymmetry is spontaneously broken for sufficiently small $\lambda_3$, and that more ground states than three occur in the frustration-free case $\lambda_3=\lambda_I$. Even more strikingly, there occur excitations with dispersion $E\propto p^3$ along the transition line \cite{Sannomiya2017b}. Our analysis thus suggests that these arise because of the phase transition to the incommensurate phase. We have checked numerically that the tricritical transition stops at the supersymmetric line. However, numerical roadblocks (in particular the presence of the $p^3$-dispersing excitations) have stopped us from further understanding of the multicritical point where the two transition lines meet. 
We plot the full phase diagram in Fig. \ref{Phase_diagram}. 


\paragraph{Conclusion}
We have introduced and analyzed a lattice Hamiltonian in the universality class of the tricritical Ising CFT and its self-dual perturbations. We found a frustration-free point with degenerate ground states, as well as lattice analogs of the supersymmetry currents. A fascinating open question is to characterize the $\lambda_c=\lambda_I$ line more thoroughly. At least three interesting phenomena occur there: the tricritical transition line terminates, there is a chiral phase transition with gapless $p^3$-dispersing modes, and the lattice supersymmetry becomes exact.

\begin{acknowledgments}
We thank H. Katsura for fruitful discussions. This work was supported by EPSRC through grant EP/N01930X.
\end{acknowledgments}

\bibliographystyle{apsrev4-1}

\bibliography{References}
\vfill\eject

\appendix

\begin{center}
\bf APPENDIX
\end{center}

\renewcommand{\theequation}{\Alph{section}\arabic{equation}}

\section{Ground states of the frustration-free model}
\label{Ising_MG_GS_proof}
In the main text it was stated that there are exactly three ground states at  $\lambda_3=\lambda_I$ (and $\lambda_c=0$). The purpose of this appendix is to prove this assertion. At this point the Hamiltonian becomes $H=\sum_j H_{j,j+1,j+2}$, where
\begin{align}
\nonumber H_{j,j+1,j+2}=&\left(\mathds{1}_j-\sigma_j^x\right) \left(\mathds{1}_{j+1,j+2}-\sigma_{j+1}^{z}\sigma_{j+2}^z\right) +\\
& \left(\mathds{1}_{j,j+1}-\sigma_{j}^{z}\sigma_{j+1}^z\right) \left(\mathds{1}_{j+2}-\sigma_{j+2}^x\right).
\label{Hjdef}
\end{align} 
As $H_{j,j+1,j+2}$ is a sum of projectors, its eigenvalues are non-negative and hence the eigenvalues of $H$ are also non-negative. Any state annihilated by $H$ is then necessarily a ground state. Conversely, all zero-energy ground states must necessarily be annihilated by each $H_{j,j+1,j+2}$ individually. 

Of the eight basis states on the three sites $j,j+1$ and $j+2$, four of them are annihilated by  $H_{j,j+1,j+2}$. The eigenstates corresponding to the zero eigenvalues are
\begin{align}
&\ket{{\cal G}^{(3)}_{j,\uparrow}}=\ket{\uparrow\uparrow\uparrow},\quad \ket{{\cal G}^{(3)}_{j,\downarrow}}=\ket{\downarrow\downarrow\downarrow},\quad
\ket{{\cal G}^{(3)}_{j,\updownarrow}}=\ket{\updownarrow\updownarrow\updownarrow}, 
\cr &\ket{{\cal G}^{(3)}_{j,s}}=\ket{\uparrow\uparrow\downarrow}+\ket{\downarrow\uparrow\uparrow}+\ket{\downarrow\uparrow\downarrow} 
-\ket{\uparrow\downarrow\uparrow}-\ket{\uparrow\downarrow\downarrow}-\ket{\downarrow\downarrow\uparrow},
\label{gs3}
\end{align}
where $\ket\updownarrow=(\ket{\uparrow}+\ket{\downarrow})/\sqrt{2}$. 
The state $\ket{{\cal G}^{(3)}_{j,\updownarrow}}$ is not orthogonal to $\ket{{\cal G}^{(3)}_{j,\uparrow}}$ or $\ket{{\cal G}^{(3)}_{j,\downarrow}}$, but these three are linearly independent and are orthogonal to $\ket{{\cal G}^{(3)}_{j,s}}$. 
Thus the three states $\ket{{\cal G}_\downarrow}$, $\ket{{\cal G}_\uparrow}$
and $\ket{{\cal G}_0}$ described in the main text are annihilated by all $H_{j,j+1,j+2}$, and so indeed are zero-energy ground states.

To determine if there are further ground states now consider the four sites $j-1,j,j+1,j+2$.
From \eqref{gs3}, it follows that such ground states can be written as a linear combination of the eight configurations $|\uparrow{\cal G}^{(3)}_{j,r}\rangle$ and $|\downarrow{\cal G}^{(3)}_{j,r}\rangle$ on these four sites,
where $r$ is one of $\uparrow,\downarrow,\updownarrow,s$. 
Since all zero-energy ground states must be annihilated by $H_{j-1,j,j+1}$ as well, they must also be linear combinations of $|{\cal G}^{(3)}_{j-1,r'}\uparrow\rangle$ and $|{\cal G}^{(3)}_{j-1,r'}\downarrow\rangle$. 
Thus we need to look for non-vanishing coefficients
$\alpha_{r,\uparrow},\alpha_{r,\downarrow},\beta_{r',\uparrow},\beta_{r',\downarrow}$ satisfying
\begin{align*}\sum_r&\left(\alpha_{r,\uparrow}\ket{\uparrow{\cal G}^{(3)}_{j,r}}+ \alpha_{r,\downarrow}
\ket{\downarrow{\cal G}^{(3)}_{j,r}}\right)\cr &\qquad = \sum_{r'}
\left(\beta_{r',\uparrow}\ket{{\cal G}^{(3)}_{j-1,r'}\uparrow}+ \beta_{r',\downarrow}  \ket{{\cal G}^{(3)}_{j-1,r'}\downarrow}\right)\ .
\end{align*}
Obviously, three solutions of this equation correspond to the states 
\begin{align}
\ket{\uparrow\uparrow\uparrow\uparrow},\quad 
\ket{\downarrow\downarrow\downarrow\downarrow}, \quad 
\ket{\updownarrow\updownarrow\updownarrow\updownarrow},
\label{gs4}
\end{align}
Any other solutions must involve non-vanishing coefficients with $r=s$ and $r'=s$. 
Comparing configurations $\ket{\uparrow\downarrow\downarrow\uparrow}$ on the two sides means $\alpha_\uparrow=\beta_{s,\uparrow}$,
and likewise comparing  $\ket{\downarrow\uparrow\uparrow\downarrow}$ 
gives $\alpha_\downarrow=\beta_{s,\downarrow}$.
However, comparing  $\ket{\uparrow\downarrow\uparrow\uparrow}$ on the two sides means $\alpha_\uparrow=-\beta_{s,\uparrow}$, and comparing $\ket{\downarrow\uparrow\downarrow\downarrow}$ 
gives $\alpha_\downarrow=-\beta_{s,\downarrow}$. These coefficients must vanish.

Thus the three states in \eqref{gs4} are the only ground states possible on four sites. This argument can then be rerun, increasing the number of sites by one each time. Thus for any number of sites greater than 3, there are exactly three zero-energy ground states.

\section{Other 4-Majorana perturbations}
In terms of Majorana fermion operators our model in the chirally invariant case can be written as
\begin{align}
H=\sum_j \left(2i\lambda_I \gamma_j\gamma_{j+1} - \lambda_3 \gamma_{j-2}\gamma_{j-1}\gamma_{j+1}\gamma_{j+2}\right)\ .
\end{align}
The latter are not the only four-Majorana terms both chirally invariant and involving nearest- and next-nearest-neighbor interactions. Indeed, Refs.\ \onlinecite{Rahmani2015a,Rahmani2015b} studied at length the effect of adding 
\begin{align}
H_R&=\sum_j \gamma_{j-1}\gamma_j\gamma_{j+1}\gamma_{j+2}\cr
&=-\sum_j \left(\sigma_{j-1}^z\sigma_{j+1}^z+\sigma_j^x\sigma_{j+1}^x\right)
\end{align}
to the Ising chain. Another such four-Majorana term is
\begin{align}
H_y&=\sum_j \left(\gamma_{j-2}\gamma_j\gamma_{j+1}\gamma_{j+2} - \gamma_{j-2}\gamma_{j-1}\gamma_j\gamma_{j+2}\right)\cr
&=\sum_j\biggl(\sigma_{j-1}^z\sigma_j^y\sigma_{j+1}^x+\sigma_{j-1}^x\sigma_{j}^y\sigma_{j+1}^z\cr
&\qquad\quad\qquad -\sigma_{j-1}^y\sigma_{j+1}^z-\sigma_{j-1}^z\sigma_{j+1}^y\biggr)\ .
\end{align}
We thus consider the Hamiltonian 
\begin{align}
H=2\lambda_IH_I+\lambda_3H_3+\lambda_R H_R +\lambda_yH_y\ .
\end{align}

\begin{figure}[t]
\includegraphics[width=1\linewidth]{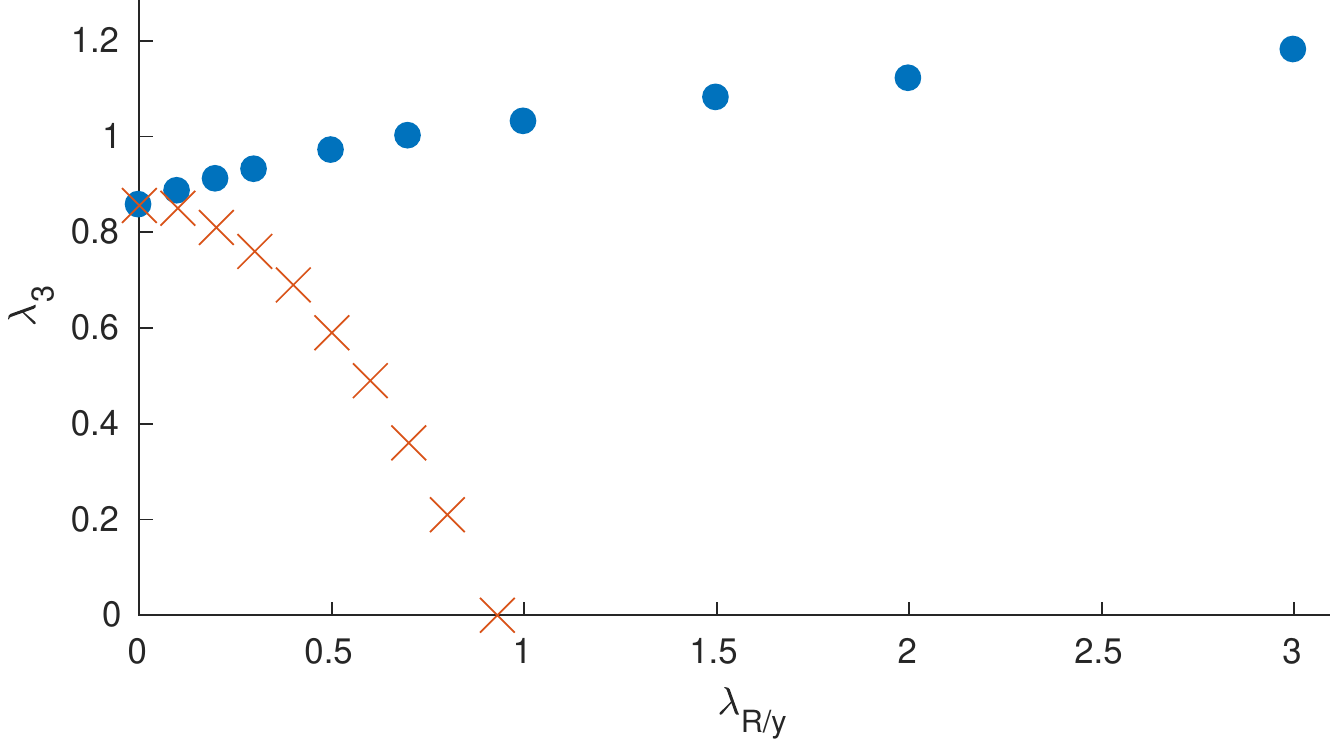}
\vspace{-0.5cm}\caption{Locations of the TCI transition in phase space for $\lambda_3$ as a function of $\lambda_R$ with $\lambda_y=0$ (blue dots), and a function of $\lambda_y$ with $\lambda_R=0$ (orange crosses).}   
\label{lR_ly}
\end{figure}

All self-dual perturbations of the Ising conformal field theory are irrelevant, with the leading irrelevant operator given by $T\overline{T}$. Thus naively, one would expect that each of the operators $H_3$, $H_R$ and $H_y$ would correspond to $T\overline{T}$, and so tuning any or all of the corresponding coefficient sufficiently will result in reaching the tricriticial point. Ref.\ \onlinecite{Rahmani2015a} found that for $\lambda_3=\lambda_y=0$, the tricritical Ising point is at $\lambda_R\approx 250\lambda_I$, a (still) mysteriously large value. In the text, we located the tricritical point for $\lambda_R=\lambda_y=0$ at $\lambda_3\approx 0.856\lambda_I$.
In Fig. \ref{lR_ly}, we locate approximately the tricritical point in the two cases $\lambda_R=0$ and $\lambda_y=0$. When both couplings are included the transition is found to be between the two cases. We thus confirm that indeed the tricritical point occurs with any or all of the perturbations, with $\lambda_y$ and $\lambda_3$ behaving qualitatively similarly, causing the transition at couplings of order 1. In keeping with the mystery of the large coupling found in Ref. \onlinecite{Rahmani2015a}, making $\lambda_R$ small and positive requires increases the other couplings further to reach the tricritical point.

\section{The incommensurate phase}

\begin{figure}[h]
\includegraphics[width=1\linewidth]{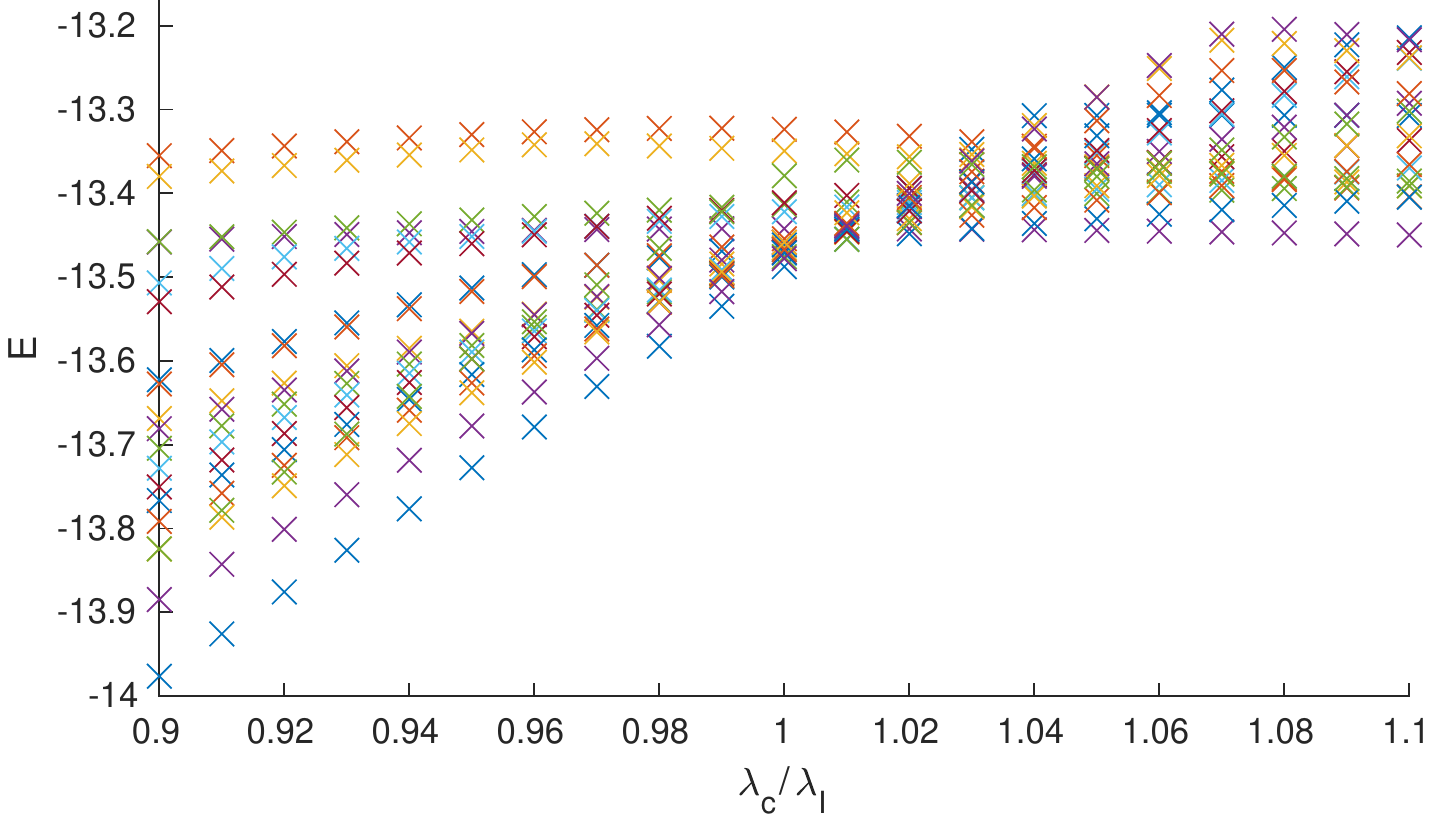}
\vspace{-0.5cm}\caption{Lowest energy levels in each momentum sector for $\lambda_3/\lambda_I=0.856$, $L=19$ and $\mathcal{F}=1$ as $\lambda_c/\lambda_I$ is taken across the chiral transition to the incommensurate phase. The Hamiltonian is scaled so that $\lambda_I+\lambda_3+\lambda_c=1$ for all values of $\lambda_c/\lambda_I$. The blue crosses (lowest for $\lambda_c/\lambda_I=1$) are in the zero momentum sector. The purple crosses (lowest for $\lambda_c/\lambda_I=1.1$) are in the $k=6\pi/19$ sector.}  
\label{Incomm}
\end{figure}

For $|\lambda_I|>\lambda_c$ the model is in an incommensurate phase. In this phase the ground state changes very rapidly as $\lambda_c/\lambda_I$ is varied. We have verified this numerically for non-zero $\lambda_3$. For $\lambda_3=0.856 \lambda_I$ the lowest energy states in the $\mathcal{F}=1$ sector are plotted against $\lambda_c/\lambda_I$ for each momentum sector with $L=19$ in Fig. \ref{Incomm}. The ground state is in the zero momentum sector up to $\lambda_c/\lambda_I=1$. After this there are many level crossings. This behaviour is reproduced all along the line and becomes more striking with larger lattice length, $L$. In the $L\to\infty$ limit an infinitesimal change in $\lambda_c/\lambda_I$ will mean the ground state is now a completely different state in a completely different momentum sector. 

The special case $\lambda_3=0$ can be solved exactly using Majorana fermions even with $\lambda_c\ne 0$. This calculation is standard, and we are sure it is presented somewhere in the literature. However, we do not know of a reference. Thus we present this calculation here, in perhaps a slightly different way than the standard Bogoliubov transformation. Throughout this we will be following the method introduced in \cite{Lieb1961} and written out explicitly for Ising in \cite{Fendley2014}.
We show that indeed there is a transition from a critical Ising phase to an incommensurate phase at the supersymmetric point. This also has the advantage of allowing for disordered couplings.


We first write down the Hamiltonian generally in terms of Majoranas, where we allow arbitrary couplings at each site:
\begin{align}
H=i\sum\limits_{j=1}^{2L} \left(2\lambda_j \gamma_j\gamma_{j+1} - \kappa_j \gamma_{j-1}\gamma_{j+1}\right)\ .
\end{align}
Here $L$ is the length of the lattice of spins, $\gamma_j$ is a Majorana obeying $\{\gamma_j,\gamma_k\}=2\delta_{jk}$, and the coefficients $\lambda_j,\kappa_j$ are chosen such that we have the Ising model (with different couplings per site) for $\kappa_j=0$ $\forall j$, and the supersymmetric model for $\lambda_j=\lambda=\kappa_j$ $\forall j$. Note that we define $\gamma_{j+2L}=\gamma_j$ and so we have periodic boundary conditions on the fermions. To get antiperiodic boundary conditions we simply change the sign of $\lambda_{2L}, \kappa_{2L}$ and $\kappa_{1}$.

We next construct raising and lowering operators. If we can find $\mu_j$ such that $\left[H,\Psi\right]=2\epsilon\Psi$, then taking any eigenstate $\ket{E}$ of $H$ with energy $E$ and applying $\Psi$ we either have $H\Psi\ket{E}=(E+2\epsilon)\Psi\ket{E}$ or $H\Psi\ket{E}=0$, hence $\Psi\ket{E}$ is either an eigenstate of the Hamiltonian with energy $E+2\epsilon$ or is a null vector. We consider the operator 
\begin{align}
\Psi=\sum\limits_{j=1}^{2L}\mu_j\gamma_j\ ,
\end{align}
where $\mu_j$ are arbitrary so far. The commutation relation
\begin{align}
\left[\gamma_j\gamma_k,\gamma_l\right]=\gamma_j\delta_{kl}-\gamma_k\delta_{jl},
\end{align}
gives
\begin{align}
\label{recursion}
&\left[H,\Psi\right]=\sum\limits_j \mu_j'\gamma_j\ , \\ \nonumber
\mu_j'=i\big(2\lambda_j\mu_{j+1}-&2\lambda_{j-1}\mu_{j-1}-\kappa_{j+1}\mu_{j+2}+\kappa_{j-1}\mu_{j-2}\big)\ .
\end{align}
Using Equation \ref{recursion} we can then find every creation operator. The ground state is the state annihilated by all operators which lower the energy and not to be annihilated by any which raise the energy.

Going to the self-dual and translation-invariant model with periodic boundary conditions, we set $\lambda_j=\lambda_I$, $\kappa_j=\lambda_c$, $\forall j$ and hence the equations are trivial to solve with the Ansatz
\begin{align}
\mu_j=\mu e^{ikj},
\end{align}
where $k=n\pi/L$, $n=-L+1,-L+2,...,-1,0,1,...,L-1,L$ to ensure $\mu_j=\mu_{j+2L}$. These indeed satisfy the recursion relations and give 
\begin{align}
2\epsilon_k=-4\sin (k)\left(\lambda_I-\lambda_c\cos (k)\right)\ .
\end{align}
We then see that $\epsilon_{-k}=-\epsilon_k$. We also note that
\begin{align}
\Psi_{-k}=\Psi_k^{\dag}, \quad \{\Psi_k,\Psi_k'\}=2L\delta_{k',-k}\ ,
\end{align}
and hence we can identify $\Psi_k$ as an annihilation operator for $0<k<\pi$ with $\Psi_{-k}$ the corresponding creation operator assuming that $\lambda_c\leq \lambda_I$. 

There is a slight subtlety if $k=0$ or $k=\pi$, as here the energies are zero and the creation and annihilation operators are the same. To resolve this we first realise that if $\Psi_{0}$ is an allowed operator with the total number of states and boundary conditions then $\Psi_{\pi}$ must be too. We therefore consider the two combinations
\begin{align}
\Psi_+=\Psi_{0}+i\Psi_{\pi}, \quad \Psi_-=\Psi_{0}-i\Psi_{\pi}\ .
\end{align}
These operators now satisfy $\{\Psi_+,\Psi_+\}=\{\Psi_-,\Psi_-\}=0$, $\{\Psi_+,\Psi_-\}=0$ and $[H,\Psi_+]=[H,\Psi_-]=0$ (note that what we have done is only sensible as $\Psi_{0}$ and $\Psi_{\pi}$ have the same eigenvalue under commutation with $H$). We now make the arbitrary choice that $\Psi_+$ is one of the creation operators (along with $-0<k<\pi$) and $\Psi_-$ is an annihilation operator. As we have now ensured that we always have $L$ independent creation operators with corresponding annihilation operators we have covered the whole space. 

For antiperiodic boundary conditions we consider the same but this time $\lambda_{2L}=-\lambda_I$ and $\kappa_{2L},\kappa_1=-\lambda_c$. Our Ansatz is then adapted to
\begin{align}
\tilde{\mu}_j=\tilde{\mu} e^{ipj},
\end{align}
where $p=q\pi/L$, $q=-L+1/2,-L+3/2,...,-1/2,1/2,...,L-3/2,L-1/2$ (so $\tilde{\mu}_j=-\tilde{\mu}_{j+2L}$). We therefore get the same formula for $\epsilon_p$ as $\epsilon_k$ above but the allowed values of $k$ differ from those of $p$.
It is then clear that the ground state remains constant for $-\lambda_I\leq \lambda_c \leq \lambda_I$ as the creation and annihilation operators stay the same for both periodic and antiperiodic boundary conditions. The low-lying excited states within each sector do change slightly, but this is due to the chiral term imposing a different Fermi velocity in the left- and right-moving directions so the underlying CFT is unchanged. For $\lambda_c>\lambda_I$ the ground state changes rapidly as $\lambda_c$ is changed as the set of creation operators also varies, signalling the incommensurate phase. As an interesting side point, note that for $\lambda_I=0$, $\lambda_c> 0$ we in fact have two Ising models, one corresponding to creation operators with $0 < k < \pi/2$ and one with $-\pi<k<-\pi/2$.

\end{document}